# Electric field control of magnetism in Si$_3$N$_4$ gated Pt/Co/Pt heterostructures

Jaianth Vijayakumar[1], David Bracher[1], Tatiana M. Savchenko[1], Michael Horisberger[2],

Frithjof Nolting[1], and C.A.F. Vaz[1],[*]

[1]*Swiss Light Source, Paul Scherrer Institut, 5232 Villigen PSI, Switzerland*

[2]*Neutron Optics and Scientific Computing group, Paul Scherrer Institut, 5232 Villigen PSI, Switzerland*

**Abstract**

In this work we show the presence of a magnetoelectric coupling in silicon-nitride gated Pt/Co/Pt heterostructures using X-ray photoemission electron microscopy (XPEEM). We observe a change in magnetic anisotropy in the form of domain wall nucleation and a change in the rate of domain wall fluctuation as a function of the applied electric field to the sample. We also observe the coexistence of in-plane and out of plane magnetization in Pt/Co/Pt heterostructures in a region around the spin reorientation transition whose formation is attributed to substrate surface roughness comparable to the film thickness; with such domain configuration, we find that the in-plane magnetization is more sensitive to the applied electric field than out of plane magnetization. Although we find an effective magnetoelectric coupling in our system, the presence of charge defects in the silicon nitride membranes hampers a systematic electrostatic control of the magnetization.

---

[*] Corresponding author. Email address: carlos.vaz@psi.ch

## I. Introduction

Multiferroics are a class of materials characterized by the presence of, and coupling between, multiple order parameters, such as between magnetic and ferroelectric order (magnetoelectric coupling) [1][2][3]. Such coupling makes it possible to control magnetism by means of applied electric fields and the ferroelectric polarization by applied magnetic fields. Although this phenomenon is known to occur in single phase compounds, such as $BiFO_3$ [4][5][6][7][8], their number is relatively small and they tend to either order at relatively low temperatures or have small magnetoelectric couplings. To overcome such limitations, artificial multiferroics, consisting of a ferromagnetic materials interfaced with a ferroelectric system such that a magnetoelectric coupling is induced at the interface, have been developed and are under intense scrutiny [1][9]. To date, the mechanisms for the control of magnetism in artificial multiferroics involve either the use of charge modulation, the creation of mechanical strain, or through interfacial exchange bias effects in the case of a multiferroic antiferromagnetic (AFM) combined with a ferromagnetic material [1][10].

Charge-mediated magnetoelectric coupling has been demonstrated in artificial multiferroic heterostructures such as $La_{1-x}Sr_xMnO_3/BaTiO_3$ [11][12], $PbZrTiO_3/La_{1-x}Sr_xMnO_3$ [13][14][15][16][17][18] and Fe/PMN-PT [19]. Charge-mediated coupling involves the modulation of the interfacial charges or movement of ions across the interface; the former shifts the Fermi level of the ferromagnetic layer resulting in a change in the density of spin up and down electrons and to a modification of the magnetic anisotropy, magnetic moment, or the magnetic ground state [20][21], while the latter involves the movement of ions towards the interface and a change in the magnetic properties by a chemical mechanism. For example, a strong modification in the interfacial anisotropy in a $Co/GdO_x$ bilayer is found to occur as a result of the movement of oxygen ions to the interface with the applied electric field [22][23]. Charge mediated coupling due to modulation of charges is preferred over coupling due to ionic movements as the movement of charge is faster, hence, more suitable for high frequency applications. However, while large charge mediated couplings can be induced at ferromagnetic/ferroelectric interfaces, the use of a ferroelectric comes with many challenges, including the need for the growth of high quality ferroelectric films with good electrical properties and smooth surfaces, since sharp and smooth interfaces are important for large magnetoelectric couplings [24][25].

To overcome these issues, the ferroelectric can be replaced by a dielectric whose growth process has been optimized. Although the permittivity of dielectrics is typically smaller than that of ferroelectrics, sizeable magnetoelectric couplings can still be found at the interface of ferromagnetic and various dielectric layers (MgO [26][27], $Al_2O_3$ [28][29], $GdO_x$ [22][23], ZnO [64][65][66]) and also in dielectric stacks [30][21]. In this work, we show the presence of a magnetoelectric coupling in silicon nitride membrane-gated Pt/Co/Pt heterostructures determined using X-ray photoemission electron microscopy (XPEEM) and magneto-optic Kerr effect (MOKE) magnetometry. The advantage of using silicon nitride is that it can be produced as a thin membrane, with good surface and insulating properties. Furthermore, with better insulating properties, higher electric fields can be applied that allow one to obtain charge modulations at the interface similar to that of high dielectric or ferroelectric materials. Pt/Co/Pt heterostructures are known to have a strong interfacial perpendicular magnetic anisotropy (PMA) [31][32][33], whose strength can be modified by electric fields [28]. To achieve a higher sensitivity of the magnetic state to an applied electric field and therefore higher susceptibilities, we consider here Pt/Co/Pt heterostructures with the Co thickness set to the spin reorientation transition (SRT), at around 6 Å, where a change in the magnetization direction occurs from out-of-plane (OOP) to in-plane (IP) with increasing Co thickness [34][35][36][37]. In addition, thermal excitations may lead to domain wall (DW) fluctuations [38][39]. Real time DW fluctuations have been observed using XPEEM [40][41], with an amplitude that was found to vary with film thickness [40] and temperature [39][42]. For example, the rate/amplitude of DW fluctuation is higher near the SRT and becomes smaller away from the SRT. In this work, by acquiring sequences of XPEEM images with time, we show that we are able to control the anisotropy of the system electrostatically in $Si_3N_4$/Pt/Co/Pt heterostructures, which is manifested in a control of the DW fluctuations, of the domain configuration, and of the SRT. By increasing the Co thickness by about 0.5 Å around the SRT, we observe the simultaneous presence of IP and OOP domains with a controllable change in the domain pattern with the applied electric field. Finally, we show that the presence of charge defects (possibly in combination with surface roughness) in the silicon nitride membrane affects negatively the electric field control of the magnetic anisotropy.

**II. Sample growth and characterization techniques**

Silicon nitride membranes were used as a substrate and also as a gate dielectric to investigate the electric field control of magnetism in the Pt/Co/Pt system. High resistivity and low stress silicon nitride membranes with thickness of 200 nm and window size of 500 × 500 $\mu m^2$ and 1 μm thick membranes with window size of 1 × 1 $mm^2$ were used. Silicon nitride has a dielectric constant of ~7.5 and a band gap of 5 eV [43] and is commonly used as an insulator in electronic devices. From atomic force microscopy, we obtain a surface roughness value of ~0.5 nm (rms) for the 200 nm membranes and ~1-2 nm (rms) for the 1 μm thick membranes. Before metal deposition, the membranes were cleaned in a piranha solution ($H_2O_2$ + $H_2SO_4$). The top electrode Cr (3 nm)/Cu (80 nm)/Cr (3 nm) and the bottom electrode Cr (3 nm)/Cu (50 nm)/Cr (3 nm) were deposited using thermal evaporation with a shadow mask. A layer of Cr is deposited before and after Cu deposition to increase the adhesion to the membrane and prevent oxidation of Cu, respectively. At this thickness, the Cu layer does not strain the membrane, which was confirmed by checking with an optical microscope, since a strained membrane can be identified from folding at the edges. The Pt (10 Å)/Co (6 or 6.5 Å)/Pt (15 Å) layers are grown after the electrode deposition using magnetron sputtering with a base pressure of $\sim 10^{-7}$ mbar. We find that the SRT for this heterostructure grown using our sputtering system occurs between 6 Å – 7 Å Co thickness. The bottom Pt layer is kept at a small value of 10 Å in order to maximize the charge screening-induced magnetoelectric effect [44].The Pt layer also protects the Co from contamination from any adsorbates on the silicon nitride membrane (deposition of Co directly on silicon nitride resulted in oxidation of Co). Finally, reference marker structures for XPEEM made of Cr were patterned using e-beam lithography. The final fabricated samples appear similar to the schematic shown in Fig. 1(a).

The XPEEM measurements performed here were carried out at Surface/Interface Microscopy (SIM) beamline at the Swiss Light Source (SLS) [45][46]. XPEEM is a spectro-microscopy technique that allows one to acquire spatially resolved spectroscopic data from the sample. In XPEEM, the sample is illuminated with monochromatic X-ray photons, resulting in the excitation of electrons from core levels to empty states in the conduction band [47]. The electron excitation process leads to the ejection of secondary electrons, whose intensity is proportional to the X-ray absorption. These secondary electrons are then accelerated toward the microscope lenses which generate a magnified image of the local X-ray absorption of the sample. In combination with the X-ray magnetic circular dichroic effect [48][49], magnetic contrast images can be acquired by using right ($C_+$) and left ($C_-$) circularly polarized light. Images with suitable photon energy (typically at the $L_3$ edge) and with $C_+$ and $C_-$ polarization are acquired separately, and by performing a pixel-wise $(C_+ - C_-)/(C_+ + C_-)$ operation, one can obtain a pure magnetic contrast image [50][51]. Similarly, a sequence of images can be obtained over a range of X-ray photon energies to yield spatially resolved spectroscopic data. XPEEM sample holders with electrical contacts to the sample [52] are used to apply electric fields *in situ*; the measurement geometry is shown in Fig. 1(a). The applied voltages are current-limited to 100 µA to avoid local sample heating. For applied voltages of 10 V on a 200 nm thick membrane, a power of around 0.4 nW/µm$^2$ and for a 1 µm thick membrane, a power of around 0.1 pW/µm$^2$ were estimated. The top electrode/surface of the sample is locally connected to the local ground and the electric field is applied from the bottom electrode; therefore, with applied positive electric fields, the Pt/Si$_3$N$_4$ interface accumulates electrons. The XPEEM, MOKE (Durham Magneto Optics NanoMOKE3) and electrical measurements were all carried out at room temperature.

## III. Results and Discussion

### III.1 Electric field control of domain wall fluctuations at the SRT with PMA

Magnetic hysteresis curves of a $Si_3N_4$ (200 nm)/Pt (10 Å)/Co (6 Å)/Pt (15 Å) heterostructure obtained using MOKE are shown in Fig. 1(b). We observe the presence of a dominant PMA [32][33][53][31] in the sample, as shown by the presence of a square loop in the OOP geometry and a hard axis loop in the IP geometry. The presence of a large coercivity and non-zero remanence in the IP hysteresis suggests the presence of an IP magnetization component. The application of electric fields results in small changes in the coercive field. Before XPEEM characterization, the sample was saturated in a magnetic field of 4 kOe out of the plane. The initial domain pattern [ Fig. 1(c)] shows the presence of small reverse domains in an otherwise uniformly perpendicularly magnetised state. From time sequences of XMCD images, we observe nucleation of reverse domains and fluctuations in the DW position in most parts of the imaged area. XMCD image sequences were acquired as a function of applied voltage at ±6, ±10, and ±15 V. At ±6 V, we find a decrease in the rate of fluctuation of the domain wall position. At +10 V we observe an increase in the DW fluctuation rate, with larger fluctuation amplitudes (~1 µm) and a much faster nucleation of magnetic domains (with an estimated DW velocity of ~0.15 µm/s). At -10 V we observe less fluctuations and no reverse domain nucleation, i.e., the DW was stable as long as the voltage was continuously applied. The control of the DW fluctuation rate confirms the presence of a magnetoelectric coupling between the silicon nitride dielectric and the Pt/Co/Pt.

To understand further the control of the DW fluctuation as a function of electric field, XPEEM image sequences were processed to quantify the rate of change in the DW fluctuation rate and amplitude [54] [55]. Figure 2(a-c) shows the processed images for different applied voltages; the pixels in red represent regions where the DW moved once or more in a twenty image sequence, while green pixels represent stable DWs; one can notice that at +10 V (Fig. 2(a)) and 0 V (Fig. 2(b)) there are more red regions than at -10 V (Fig. 2(c)). The zoomed-in view of the domain inside the orange box (insets), shows more clearly that at -10 V the domain is very stable, with almost no fluctuations, while at + 10 V a change in shape occurs together with an increase in the DW fluctuation rate. To calculate the displacement of the DW in the image sequence, we calculate the relative change of the perimeter of the same domain wall with respect to the first image in the sequence; a positive value indicates nucleation and increase in domain size while a negative value corresponds to domain shrinking. The perimeter is calculated to first approximation by summing the number of pixels in the image, since these processed DW images have intensity '1' in the DW position and '0' elsewhere, to provide information about how much the domain shape has changed from the initial state and by how much per each image in the sequence. The DW displacement for the domains enclosed in the white box in Fig. 2(a), is shown in Fig. 2(d) for 0 V, +15 V and -15 V; the horizontal black lines correspond to the DW perimeter of the first image at the respective voltage. We find that the DW perimeter at 0 V and 15 V fluctuates strongly and does not return to its initial value, indicating that the DW perimeter or the size of the domain is increasing with time. For -15 V, we observe small fluctuations, predominantly in the vicinity of the initial state, while the decrease in the perimeter of the DW from the initial state indicates that the domains are shrinking with time. Fig. 2(e) shows the average change in DW perimeter acquired as a function of the applied voltage, where it is clear that positive electric fields favor nucleation and negative electric fields favor reverse nucleation. The video of the DW fluctuations as a function of electric field can be found in the supplementary information (S1- "Multimedia view"). A single XMCD image in the sequence is acquired in 20 s and from the DW fluctuation rate, one can estimate the energy barrier ($E_b$) associated with the DW fluctuations using an Arrhenius equation $f = f_0 \exp\{-E_b/k_B T\}$, where $f_0$ is an attempt frequency (~1.9 x 10$^9$ Hz) [56][57], which is defined as the number of fluctuations a spin undergoes before it flips, $k_B$ and $T$ are the Boltzmann constant and temperature, respectively, and the experimental frequency $f$ is obtained from the same DW at 0, ±10 V (±50 MV/m). Using these values, we obtain $E_b \sim 0.61$ eV for 0, +10 V, and $E_b \sim 0.68$ eV for -10 V, which corresponds to a change of about 10 % in the energy barrier height. The equivalent change in the Fermi energy can be calculated from the charge modulation at the interface, $\Delta Q = V \varepsilon_r \varepsilon_o A/d$, where $\varepsilon_r$ is the relative permittivity, $\varepsilon_o$ is the permittivity in vacuum, $A$ (500 × 500 μm$^2$) is the area and $d$ is the silicon nitride thickness (200 nm), from which we estimate $\Delta Q \sim \pm 2.07$ x 10$^{12}$ $e$/cm$^2$ or $\Delta Q \sim \pm 0.9$ x 10$^{-3}$ $e$/Co atom. The latter could result in a change of Fermi level ($E_f = \Delta N/DOS \times d_{Co}$) by 0.2-0.7 meV, for a thickness $d_{Co}$ in the range from 1-3 monolayers, where $DOS$ is the density of states of Co at the Fermi level, ~*17.26* electrons/(atoms Ry spin) [58]. A similar value was obtained by Maruyama *et al*. [21], which was enough to modify the anisotropy of a few monolayers (1-3) Fe layer using MgO as gate dielectric. Therefore, in Pt/Co(6 Å)/Pt with PMA, at zero and positive electric field, the anisotropy energy is reduced, favoring the breaking down of the saturated domain into a multidomain state, while a negative electric field increases the anisotropy energy and favors a monodomain state, a result similar to that reported by Mantel *et al*. [30]. We note that the Pt layer is likely to be magnetically polarized through the proximity to the Co layer and that a change in magnetic moment with the electric field may occur [59][60][61][62][63], which may also contribute to the change in the magnetic energy.

In order to rule out chemical modification of the Co state with the electric field, we measured the X-ray absorption spectra (XAS) of Co under the applied voltages, shown in Fig. 2(f). We find that, while we can rule out large chemical modifications in Co as a function of electric field that could result from ion diffusion from the silicon nitride, small modifications in the spectra do occur, with the Co $L_3$ peak position shifting to lower energies and and becoming narrower as the voltage changes from -20 V to 0 V to +20 V. The trend is consistent with a shift in the Fermi energy position and with the larger magnetic changes found between 0 and +10 V as compared between 0 and -10 V, Fig. 2 (a-c).

**III.2. Electric field control of the magnetic state at the SRT with coexisting IP and OOP domains**

An increase in the Co thickness from 6 Å to 6.5 Å resulted in the presence of both IP and OOP magnetization components, as seen in the MOKE data shown in the Fig. 3(a), where we find a square hysteresis loop for both IP and OOP directions of the applied magnetic field. We also see the presence of a uniaxial in-plane anisotropy from polar plots of the IP coercivity, as shown in Fig. 3(b), hence we cannot exclude the possibility that the easy axis is canted to a certain OOP angle. Samples with 6.5 Å Co thickness were grown simultaneously on both 1 μm and on 200 nm thick silicon nitride membranes. From XPEEM we find similar domain patterns for the films grown on both membrane thicknesses, even though the surface roughness is different by about 50%; here we show the results for $Si_3N_4$ (1 μm)/Pt (10 Å)/Co (6.5 Å)/Pt (15 Å). Figure 3(c) shows the variation of the coercivity of the IP hysteresis loop as a function of electric field, applied in the field sequence described by the blue squares. We find that, irrespective of the particular field sign and amplitude, the coercive field value decreases continuously, with some fluctuations. The drop in the coercivity indicates that the IP easy axis is changing towards a hard axis. After 50 voltage cycles the coercivity changed from 10 Oe to 1 Oe (a change of ~90%), reaching a hard axis behavior, and further application of voltage excitations leads to no further changes in the magnetic response; the initial and the final IP hysteresis loop after applying 50 cycles of voltages is shown in Fig. 3(e). During MOKE characterization we simultaneously imaged the surface of the membrane using the Nano-MOKE microscope and found no detectable straining of the membrane with the applied electric field. We also find that the coercivity does not change when there is no electric field applied, as shown in Fig. 3(c) as red circles, indicating that the change in coercivity with electric field results from charge modulation at the interface. For the OOP hysteresis loops, the changes with the applied voltage were small, with a change in coercivity of less than 2%, as shown in Fig. 3(d). Therefore, based on the MOKE characterization, a sample with both IP and OOP magnetization components can be manipulated by the electric field; however, we find that IP magnetization is more sensitive than the OOP magnetization, which we relate to the sensitivity of the magnetic state to the perpendicular magnetic anisotropy at this Co thickness in combination with the fact that, considering the roughness of the membrane substrate, it is possible that the 1 nm thick Pt layer is not continuous, such that some regions of the Co layer may be in contact with the membrane; these Co regions may have weaker perpendicular magnetic anisotropy (more in-plane magnetization component) and see a stronger effect from the electric field in the dielectric substrate. Similar irreversible changes in the coercivity was previously observed due to interfacial oxide formation or direct heating effect originating from the gate oxide.

To understand the magnetic domain structure, the sample was first demagnetized using an AC magnetic field of about 15 mT and was imaged with XPEEM before applying any electric fields, as shown in Fig. 4(a). We observe irregular shaped domain patterns with a lateral dimension of the order of 1 µm. To verify the possible presence of IP and OOP magnetization, the sample was rotated azimuthally with respect to the X-ray beam direction. The XMCD effect is a magnetization vector dependent absorption process proportional to the cosine of the angle between the X-ray propagation vector and the magnetization; the incidence angle of X-rays in XPEEM is 16º, which leads to a higher sensitivity to the IP magnetization [Fig. 1(a)]. When the sample is rotated, the intensity of absorption of IP domains changes but not that of the OOP domains. Images taken at 0º and 90º show regions where the magnetic contrast changes, corresponding to IP domains, and regions where the magnetic contrast remains constant, corresponding to OOP domains. The latter are colored in magenta in Fig. 4(a) and for this particular state, it represents about 1% of the overall area imaged. This observation confirms the coexistence of IP and OOP domains, which we attribute to the relatively large surface roughness of silicon nitride membrane, comparable to the thickness of the Co layer, and also possibly to a slight non-uniformity in the thickness of Co, which is around the SRT. To learn more about the effect of the electric field in such a domain configuration, XMCD images as a function of the applied voltage were acquired and are shown in Fig. 4(b-m). We observe that the multidomain state gradually breaks down into smaller domains with increasing applied voltage and eventually the domain size becomes smaller than the microscope resolution. We observe the breaking down of domains to very small domains in both directions of the applied voltages which is agreement with the MOKE data, which shows that the coercivity reduces with applied electric field. In the image sequence shown in Fig. 4(b-m), one can notice, for positive voltages, that the transition from larger domain configuration to a very small domain configuration occurs at 9 V while for negative voltages it occurs at - 6.5 V. Upon removal of the electric field we observe that the initial larger domain configuration reappears [Fig. 4(m)], and we find that there are some small, but observable changes in the IP domains while the OOP domains marked in blue in Fig. 4(a) remain unchanged. This result agrees with the MOKE data, where we saw changes only in the IP anisotropy. For the charge carrier modulation and change in Fermi energy, we expect a values lower by a factor of 10 compared to the 100 nm thick membranes. It is unclear why we observe similar changes for both directions of the applied electric field. Interface roughness could play a role, since the roughness of the substrate is comparable to the film thickness, and in fact surface-induced changes in spin concentration has been demonstrated in Fe monolayers [67] and roughness-induced band gap modification has also been discussed in oxides [68].

Similar measurements were carried out in other Pt/Co/Pt heterostructures and we observe that most of the samples showed irreversible or random changes in domain configuration or coercivity with the applied electric field. To better understand this behavior, we characterized electrically the silicon nitride membranes by measuring the capacitance as a function of frequency using an LCR impedance meter; the results are shown in Fig. 5, where we find that the capacitance reduces with increasing frequency by almost a factor of 4 from 100 Hz to 100 kHz and to be compared to the ideal, frequency-independent capacitance of 86 pF (value which agrees well with that measured at 100 MHz, ~90 pF). At room temperature, the relative permittivity of silicon nitride does not change with frequency [69], hence, we attribute this effect to the presence charge defects/traps in the silicon nitride membranes. Even though silicon nitride is considered to be a good insulating material, the electrical and field effect properties can be affected by the deposition method, stoichiometry and the presence of oxygen used to produce low stress membranes. Such intrinsic defects result in the formation of charge traps [70][71][72][73][74] and efforts are being made to reduce/control the charge trap density [75][76]. Since the mobility of carrier charges such as electrons is independent of the measuring frequency and we measure a higher capacitance at lower frequency, it is likely that we are also moving ions or possible vacancies in the $Si_3N_4$ with applied electric field along with the electrons, as ionic mobility with electric field is lower than electron mobility. Therefore, when an electric field is applied, we accumulate charges and/or ions at the interface and in turn change the interfacial anisotropy. Upon removal of the electric field, a number of charges remain trapped in the interface, pinning the Fermi level and the anisotropy in the modified state; further application of electric field leads to further accumulation of trapped charges at the interface and a new modification of the interfacial anisotropy in an irreversible manner (within the time of measurement), which can be related to our observations in MOKE and XPEEM. Therefore, for a systematic control of magnetization, the quality of the silicon nitride, especially the stoichiometry, should be taken into account for gated devices. In fact, silicon nitride was considered for electrically read-only and flash memory devices [77], and also used as a charge trapping device to store information [78], suggesting that it may be possible to combine magnetoelectric coupling in silicon nitride gated devices and its charge trapping effects for future MRAM or memory devices. With defect free membranes, we expect a better magnetoelectric coupling which can be used for high frequency characterization and applications.

**IV. Conclusions**

In conclusion, we have demonstrated the presence of a magnetoelectric coupling in silicon nitride gated Pt/Co/Pt heterostructures using XPEEM and MOKE. From XPEEM we find that we can control the DW fluctuation rate and DW patterns using an electric field; in both cases we change the anisotropy of the system with an applied electric field. We also demonstrate the coexistence of both IP and OOP magnetization components which is attributed to surface roughness; in such systems, the IP magnetization is found to be more sensitive to the applied electric field and the control in domain wall pattern is systematic. Due to the presence of charge defects in the silicon nitride membranes that acts as charge traps, the control of magnetic state with applied electric field is found to be not fully reversible. Therefore, it is important that good quality silicon nitride membranes are used for gated devices. With optimized growth process, silicon nitride can be a potential candidate as gate dielectric in magnetoelectrically coupled devices for future electric field control MRAM or storage devices.

**Acknowledgments**

This project is funded by Swiss National Science Foundation (SNF) (Grant no. 200021_153540). Tatiana M. Savchenko is funded by SNF (Grant no. 200021_160186) and David Bracher is funded by Swiss Nanoscience Institute (SNI) (Grant no. SNI P1502). The authors wish to thank Vitaliy Guzenko from the Laboratory of Micro and Nanotechnology for his support with the e-beam lithography and Marcos Gaspar for the help with the capacitance measurements. Magnetooptical Kerr effect measurements were carried out using the Durham Magneto Optics NanoMOKE3® system of the Laboratory for Mesoscopic Systems, ETH Zurich, Switzerland and the Laboratory for Multiscale Materials Experiments, Paul Scherrer Institute, Switzerland. Part of this work was performed at the Surface/Interface: Microscopy (SIM) beamline of the Swiss Light Source (SLS), Paul Scherrer Institut (PSI), Villigen, Switzerland.

Supplementary information:

1. Videos  -10V, 0V, +10V

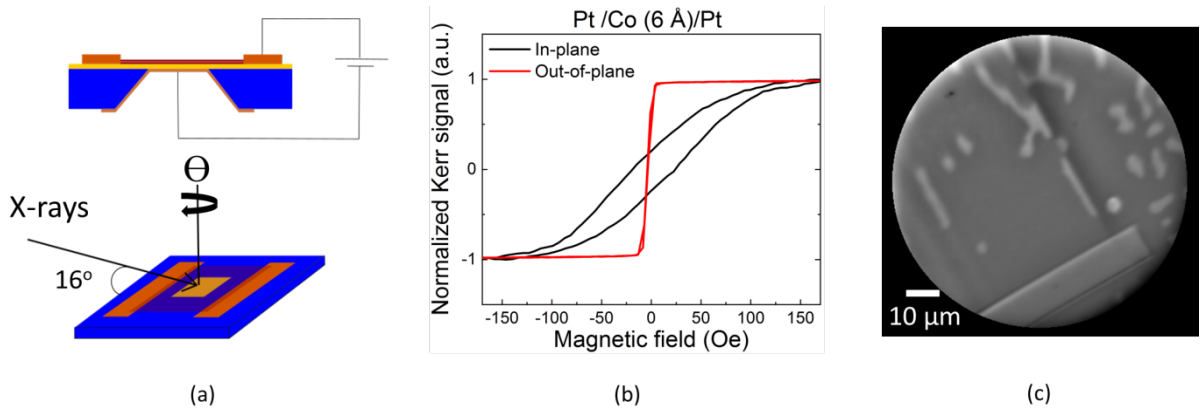

*Figure 1: (a) Schematic of the sample structure. The blue region corresponds to Si, yellow to the Si₃N₄ membrane, electrodes are shown in brown and the tri-layer Pt/Co/Pt is present between the two top electrodes; the measurement geometry in XPEEM is shown at the bottom. (b) Polar and longitudinal magneto-optic Kerr signal vs magnetic field for Pt/Co (6 Å)/Pt applied IP and OOP to the sample. (c) Magnetic contrast XPEEM image taken at the Co $L_3$ edge.*

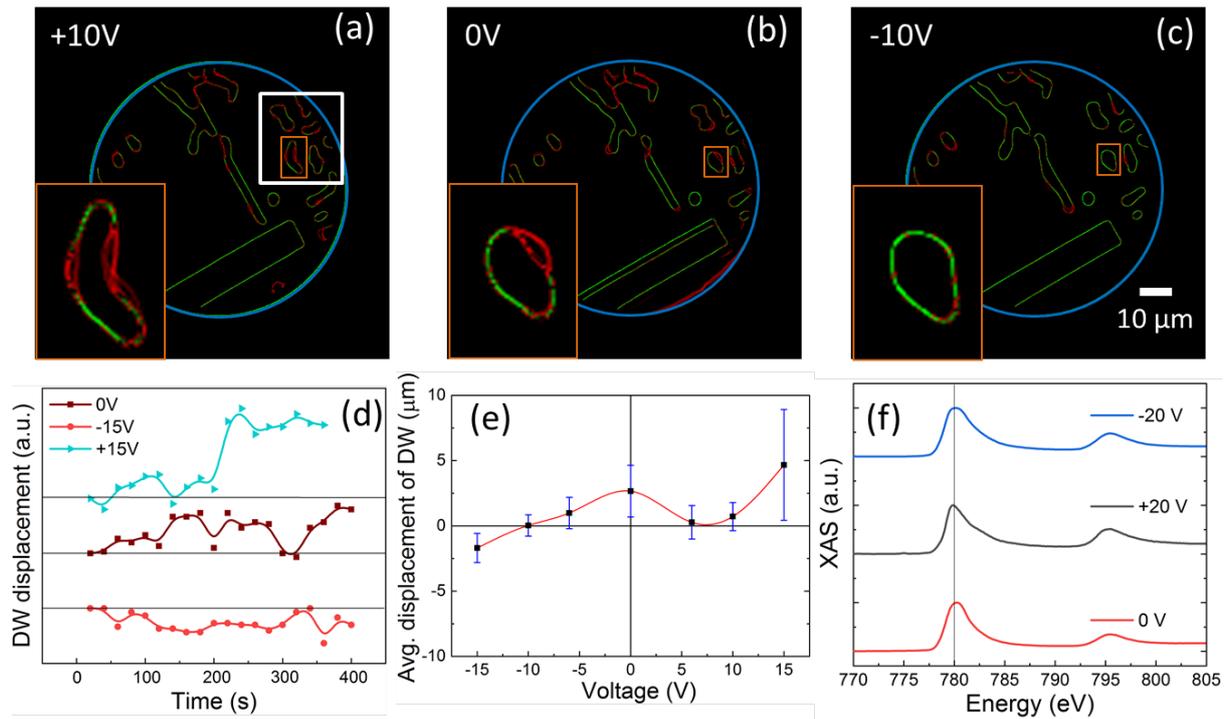

*Figure 2: (a-c) Extracted domain wall summed over all images as a function of applied voltage. The region marked in red represent regions where the DW moved once or more and green region represents stable DWs. The inset is a zoom-in of the magnetic domain in the orange box. (d) Domain wall displacement with time for different voltages; black horizontal line represents the perimeter of the DW at the first image. (e) Averaged domain wall displacement as a function of voltage. (f) XAS of Co across the $L_{2,3}$ edge under different applied voltages.*

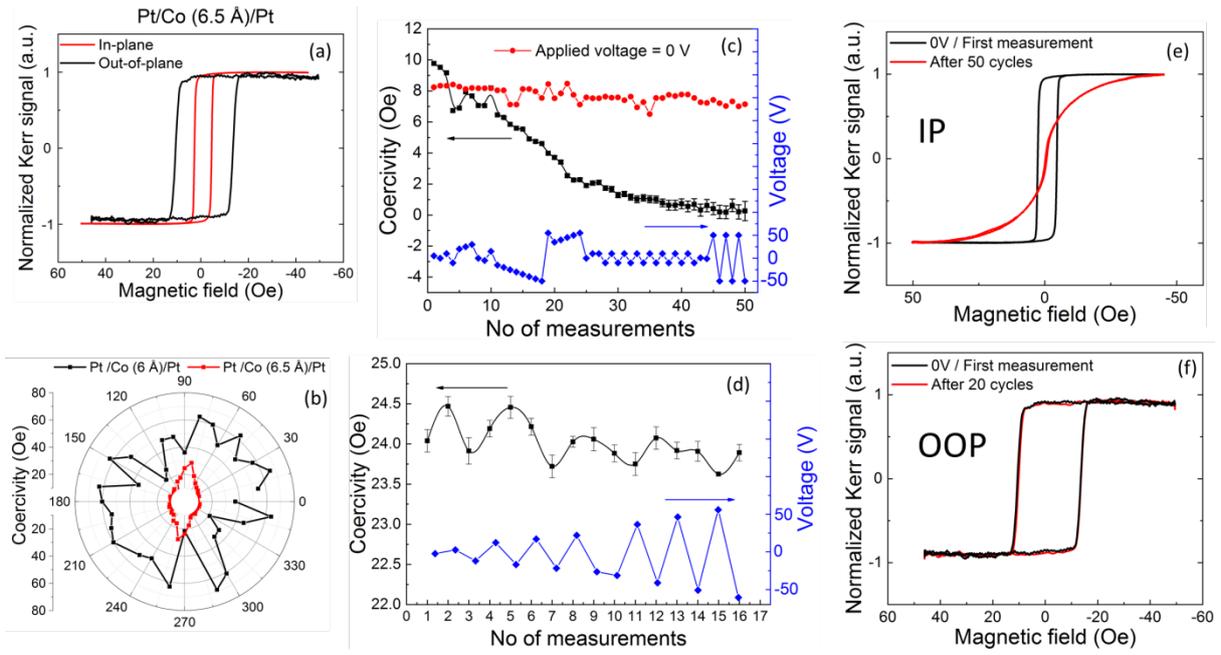

*Figure 3: (a) MOKE characterization of IP and OOP geometry of Pt/Co (6.5 Å)/Pt. (b) Polar plot of the IP coercivity of Pt/Co (6 Å)/Pt and Pt/Co (6.5 Å)/Pt. (c,d) Change in IP and OOP coercivity as a function of electric field pulse sequence, respectively (lines are guides to the eye). (e) IP Hysteresis loops before and after the application of 50 voltage cycles shown in Fig. 3(c). (f) OOP Hysteresis loops before and after the application of 16 voltage cycles shown in Fig. 3(d).*

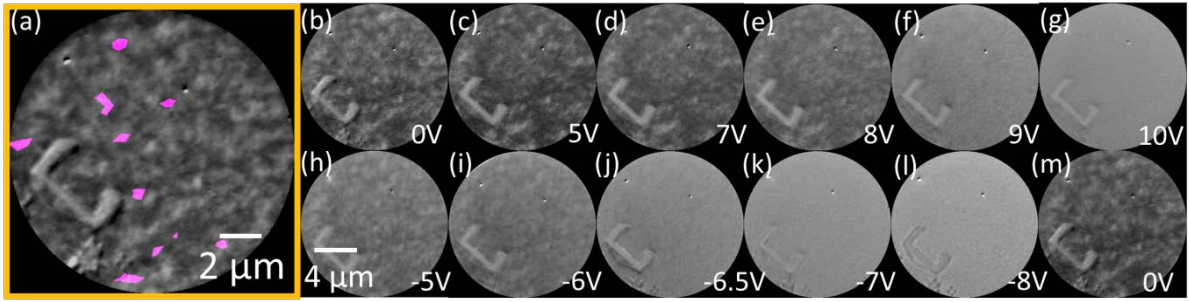

*Figure 4: (a) Domain configuration of Pt/Co (6.5 Å)/Pt, showing the presence of both IP and OOP domains (the latter marked in magenta). (b-m) Domain pattern as a function of applied voltage.*

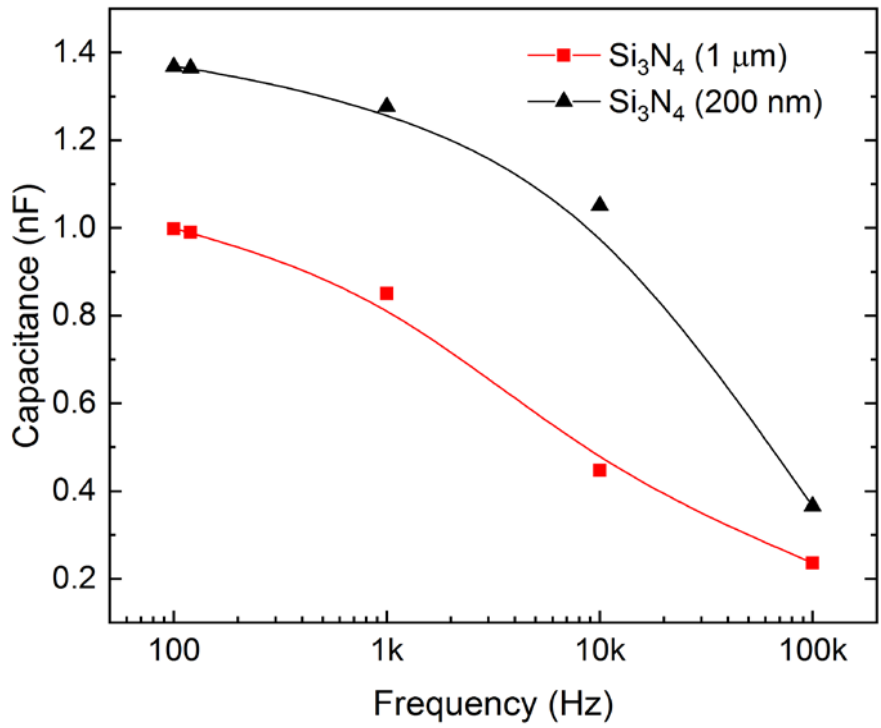

*Figure 5 : Capacitance as a function of frequency for the silicon nitride membranes (lines are guides to the eye).*

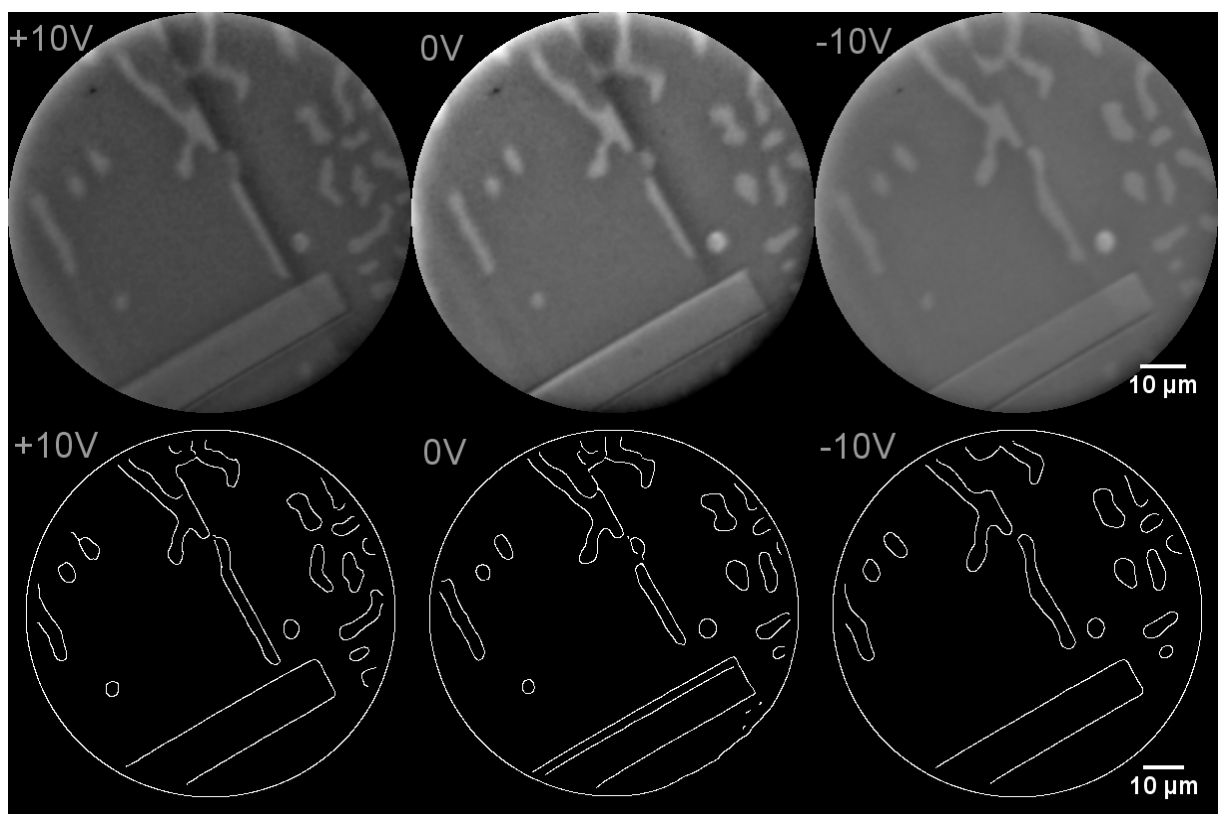

*Supplement information (S1): Domain Wall Fluctuation video "(Multimedia view)"*